%% file: main.tex
\documentclass{article}
\usepackage[T1]{fontenc} 
\usepackage[utf8]{inputenc} 
\usepackage{amsmath,cite,url}
\usepackage{graphicx}

\usepackage{subcaption}

\usepackage{amssymb}
\usepackage{pifont}

\usepackage{multirow}
\usepackage[symbol]{footmisc}
\usepackage{booktabs}
\usepackage[table,xcdraw]{xcolor}
\usepackage{import}
\usepackage{tikz}
\usepackage{ismir} 
\usepackage{enumitem}
\usepackage{array}

\newcommand{\redcross}{\textcolor{red}{\ding{55}}} 
\newcommand{\greencheck}{\textcolor{green}{\ding{51}}} 

\newcommand{\mAR}{\text{mAR}}
\newcommand{\LL}{\mathcal{L}}

\definecolor{lightgray}{RGB}{230,230,230}

\usepackage{dsfont}

\newcommand{\norm}[2]{\| #2 \|_{_{#1}}}

\newcommand{\alain}[1]{{#1}}

\usepackage{lineno}

\title{SLAP: Siamese Language-Audio Pretraining\\without negative samples for Music Understanding}

\multauthor
{Julien Guinot$^{*,1,2}$ \hspace{1cm} Alain Riou$^{*,3}$ \hspace{1cm} Elio Quinton$^2$ \hspace{1cm} György Fazekas$^1$} { 
$^1$ Centre for Digital Music, Queen Mary University of London, U.K.\\
$^2$ Music \& Audio Machine Learning Lab, Universal Music Group, London, U.K.\\
$^3$ LTCI, Télécom-Paris, Institut Polytechnique de Paris, France \\
{\small $^*$Equal contribution, correspondence to {\tt j.guinot@qmul.ac.uk}}
}

\def\authorname{J. Guinot, A. Riou, E. Quinton, and G. Fazekas.}

\usepackage[bookmarks=false,pdfauthor={\authorname},pdfsubject={\papersubject},hidelinks]{hyperref}

\hypersetup{
    colorlinks=true,
    linkcolor=blue,
    citecolor=blue,
    filecolor=magenta,      
    urlcolor=blue,
    pdftitle={Overleaf Example},
    pdfpagemode=FullScreen,
    }

\begin{document}

\maketitle

\begin{abstract}


Joint embedding spaces have significantly advanced music understanding and generation by linking text and audio through multimodal contrastive learning.
However, these approaches face large memory requirement limitations due to relying on large batch sizes to effectively utilize negative samples. Further, multimodal joint embedding spaces suffer from a modality gap wherein embeddings from different modalities lie in different manifolds of the embedding space.

To address these challenges, we propose Siamese Language-Audio Pretraining (SLAP), a novel multimodal pretraining framework that allows learning powerful representations without negative samples. SLAP adapts the Bootstrap Your Own Latent (BYOL) paradigm for multimodal audio-text training, promoting scalability in training multimodal embedding spaces.

We illustrate the ability of our model to learn meaningful relationships between music and text --- specifically, we show that SLAP outperforms CLAP on tasks such as text-music retrieval and zero-shot classification. We also observe competitive downstream performance on several MIR tasks, including with larger or supervised models (genre and instrument classification, auto-tagging).

Additionally, our approach has attractive properties, such as a quantifiably reduced modality gap and improved robustness to batch size variations on retrieval performance.
Finally, its novel formulation unlocks large-scale training on a single GPU through gradient accumulation. 

\end{abstract}


\section{Introduction}


Joint embedding spaces for text and audio have been foundational in recent developments in music understanding and generation.
Such spaces are typically learned via Multimodal Contrastive Learning (MCL), which optimizes
a pair of encoders for maximal similarity between positive pairs, while minimizing similarity for negative pairs~\cite{clip,elizalde2023clap}.

Though widely successful, some drawbacks have been identified with this method. Recent trends for joint multimodal embedding spaces have emphasized fine-grained understanding between individual text tokens and individual timesteps of music \cite{wu2024collap, komatsu2025aligned} and need for text augmentation strategies to alleviate the lack of large scale datasets \cite{yuan2024t,manco2024augment}. A modality gap emerging from the contrastive approach has been observed, where modalities lie in separate manifolds of the embedding space \cite{Fahim2024, liang2022mind}. This leads to ``joint'' representations not being truly joint, potentially harming performance. Importantly, contrastive approaches face an inherent scalability issue. Their formulation makes them 1) dependent on batch size for representation quality and 2) more difficult to scale than masked modelling approaches due to the formulation of the contrastive loss \cite{pham2023combined}. These drawbacks pose issues for foundation models, which by usage should be scalable for adaptation on large-scale datasets. Further, text-music spaces are a many-to-many space. I.e., there is often no one single corresponding appropriate caption for a piece of music, and \textit{vice-versa}. 


Inspired by recent advances in Self-Supervised Learning for images and general audio~\cite{grill2020bootstrap,niizumi2021byol},
we propose \textbf{SLAP}, \textbf{S}iamese \textbf{L}anguage \textbf{A}udio \textbf{P}retraining.
We adapt Bootstrap Your Own Latent (BYOL) \cite{grill2020bootstrap} as a joint-embedding pretraining approach without negative samples to multimodal text-audio pretraining.
We show that SLAP alleviates both
the scalability issues
and the modality gap inherent to MCL. Our contributions are:



\begin{enumerate}
\item We introduce a scalable, hyperparameter-robust approach to language-audio pretraining which does not require negative pairs to learn strong representations.
\item We outperform comparable contrastive models on retrieval and downstream probing.
\item We show that our approach significantly decreases the modality gap between audio and text embeddings compared to contrastive approaches.
\item Our approach enables larger batch sizes via gradient accumulation, which was previously inaccessible due to the formulation of the contrastive loss.
\end{enumerate}


To facilitate further research in this direction, we make our code available.\footnote{\href{https://github.com/Pliploop/SLAP?tab=readme-ov-file}{\tt https://github.com/Pliploop/SLAP}}

\section{Background}

\subsection{Multimodal Contrastive Learning}

In contrastive learning, models learn representations by maximizing the similarity between two \emph{views} of an input while minimizing its similarity with \emph{negative samples}, typically by optimizing an InfoNCE loss~\cite{chen2020simple}.
While these views are typically crafted by randomly applying transforms to the input data points~\cite{chen2020simple,spijkervet2021contrastive}, Multimodal Contrastive Learning (MCL) instead relies on paired data from different modalities, such as text-image \cite{clip}, text-video~\cite{VATT}, audio-image~\cite{L3}, text-music \cite{manco2022contrastive,mulan}, and text-audio \cite{elizalde2023clap, zhu2024cacophony}.
Contrastive learning spanning more than two modalities has also been investigated~\cite{guzhov2022audioclip,girdhar2023imagebind,GRAM}.

MCL has been leveraged to great extent in multiple modalities for tasks such as understanding \cite{clip}, retrieval \cite{manco2022contrastive}, and generation \cite{ramesh2022hierarchical}, with many multimodal foundation models emerging from the approach \cite{alayrac2022flamingo,AudioFlamingo}.

MCL, specifically text-audio and text-music contrastive learning, has also been widely adopted in the field of audio and music. Notably, CLAP \cite{elizalde2023clap} and subsequent developments building upon it have demonstrated high audio understanding performance~\cite{yuan2024t,wu2024collap,zhu2024cacophony,ghosh2025reclap}.
Learned audio-text representations enable effective text-audio and audio-text retrieval by leveraging similarity metrics between modalities. Beyond catalogue navigation, these capabilities support retrieval-augmented audio captioning \cite{li2025drcap, ghosh2024recap}.
In music-text representation learning, MusCALL \cite{manco2022contrastive} and MuLan \cite{mulan} have both achieved strong performance on downstream tasks such as text-music retrieval, music classification, and zero-shot classification.
Finally, joint embedding text-audio spaces have also been leveraged to condition audio and music generative models \cite{evans2024fast,evans2024long,diffariff,liu2023audioldm,agostinelli2023musiclm}.



\subsection{Limitations of contrastive learning}\label{sec:limits}

Despite its widespread success, contrastive learning faces notable limitations, typically related to its reliance on negative samples.
First, ensuring a diverse and representative set of negative samples is crucial for stable training, which usually necessitates a large batch size $B$ in practice.
However, computing the InfoNCE loss requires storing a $B \times B$ matrix of pairwise similarities on a single GPU, which limits scalability.
SigLIP \cite{SigLIP} addresses this by replacing the softmax in the criterion with a sigmoid, making distributed training more tractable, though it still demands costly inter-device communication.
In addition, contrastive learning implicitly assumes a uniform prior for the data distribution, which is detrimental in long-tailed scenarios \cite{Assran2023}.

MCL, more specifically, also suffers from an issue often referred to as the \textit{modality gap}, where embeddings from different modalities occupy non-overlapping cone manifolds in the joint latent space~\cite{liang2022mind}.
\cite{Shi2023} observe a relationship between this gap and the initial model weights and loss temperature, while recent findings suggest this problem could be intrinsic to the loss formulation itself~\cite{Fahim2024}.


\begin{figure*}[t]
    \centering
    \includegraphics[width=1\textwidth]{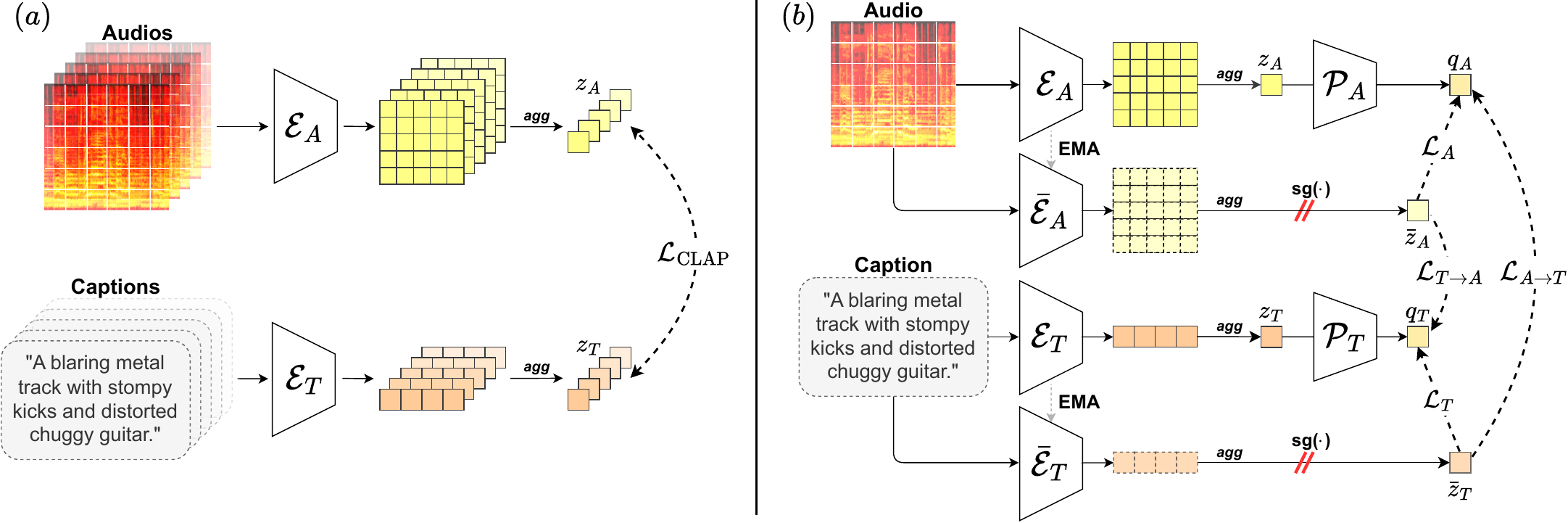}
    \caption{
        $(a)$ \textbf{CLAP:} The multimodal contrastive loss $\LL_{\text{CLAP}}$ is optimized between batches of audio/text pairs.
        $(b)$ \textbf{SLAP:} For each audio/text pair, we compute a query $q$ and a target $\bar{z}$ for each modality, then optimize both the intermodal losses $\LL_{A \to T}$ and $\LL_{T \to A}$ and the intramodal losses $\LL_A$ and $\LL_T$ between the queries and the targets.
        As indicated by the stop-gradient operator $\text{sg}$, the gradient flows only through the context branches, while the target encoders are updated via EMA.
    }
    \label{fig:model}
\end{figure*}


\subsection{Self-supervised learning without negative samples}

Recent advances in Self-Supervised Learning (SSL) have demonstrated that contrastive learning can be outperformed by alternative approaches that do not rely on negative samples, at least in unimodal settings \cite{chen2021exploring,grill2020bootstrap,assranSelfSupervisedLearningImages2023}.

In particular, BYOL~\cite{grill2020bootstrap} leverages an asymmetric architecture composed of a \emph{context} encoder and a \emph{target} encoder. Two augmented views are passed through their respective encoder, and a \emph{predictor} network maps the context output to match the target's. Crucially, the target encoder is not trained via gradient descent but updated as an Exponential Moving Average (EMA) of the context encoder, which effectively avoids representation collapse~\cite{Tian2021}.


This approach, as well as
subsequent works
relying on Transformer-based architectures, do not suffer from the conceptual limitations described in Section~\ref{sec:limits}, and achieve state-of-the-art performance in image~\cite{DINOv2,assranSelfSupervisedLearningImages2023,baevskiData2vecGeneralFramework2022} and audio~\cite{niizumi2021byol,ATST,niizumiMaskedModelingDuo2023,MATPAC} representation learning.
However, the reliance on EMA requires the context and target encoders to share the same architecture, hindering the direct application of these methods to multimodal scenarios.

In this work, we draw inspiration from these ideas to propose a novel multimodal SSL framework that eliminates the need for negative samples while addressing the architectural limitations imposed by EMA-based methods.

\section{Siamese Language-Audio Pretraining}


The training pipeline of SLAP is depicted in Figure~\ref{fig:model}. Consider an audio encoder $\mathcal{E}_A$ mapping samples $x_A$ of length $T_A$ from the audio space to a latent audio representation $z_a$  so that $\mathcal{E}_A : x_A \in \mathbb{R}^{T_A} \mapsto z_A \in \mathbb{R}^d$ and a text encoder mapping $N$ tokens to a tex latent space $\mathcal{E}_T : x_t \in \mathbb{N}^{N} \mapsto z_t \in \mathbb{R}^d$.

Audio and text encoders project inputs into respective modalities' latent spaces. For each modality, we
define a 
\emph{context encoder} ($\mathcal{E}$) and an Exponential Moving Average (EMA)-updated \emph{target encoder} $\Bar{\mathcal{E}}$ with EMA rate $\tau$:
\begin{equation}
    \Bar{\mathcal{E}} = \tau \Bar{\mathcal{E}} + (1-\tau) \mathcal{E}.
\end{equation}

For each modality, we also incorporate a \emph{predictor} $\mathcal{P}$,
which has been shown to be necessary to prevent trivial solutions and collapse
\cite{grill2020bootstrap,Tian2021}.
The predictor of each branch processes the output $z$ of the respective context encoder and aims to predict the output $\Bar{z}$ of the target branch of both modalities.
We notate $q$ the output of the predictor\alain{:}

\begin{equation}
    q_A = \mathcal{P}_{A}(z_A), \qquad
    q_T = \mathcal{P}_{T}(z_T).
\end{equation}

The model is trained by minimizing \emph{intermodal losses}, defined as the cosine distance between the queries $q$ and targets $\bar{z}$ from different modalities:
\begin{equation}
    \mathcal{L}_{A \rightarrow T} = 1 - \frac{q_A \cdot \Bar{z}_T}{\norm{}{q_A} \norm{}{\Bar{z}_T}}, \;
    \mathcal{L}_{T \rightarrow A} = 1 - \frac{q_T \cdot \Bar{z}_A}{\norm{}{q_T} \norm{}{\Bar{z}_A}},
\end{equation}
where $\norm{}{\cdot}$ denotes the $L_2$ norm.

Additionally, we introduce \emph{intramodal losses} between the queries $q$ and targets $\bar{z}$ within each modality, which we find to empirically yield stronger retrieval performance than using intermodal losses alone (see Section \ref{subsection: role of intermodal and intramodal loss terms}):

\begin{equation}
    \mathcal{L}_{A} =  1 - \frac{q_A \cdot \Bar{z}_A}{\norm{}{q_A} \norm{}{\Bar{z}_A}}, \quad
    \mathcal{L}_{T} =  1 - \frac{q_T \cdot \Bar{z}_T}{\norm{}{q_T} \norm{}{\Bar{z}_T}}.
\end{equation}

The final loss is a combination of intramodal and intermodal losses with a weighing term $\lambda \in [0, 1]$:

\begin{equation}
    \mathcal{L} = \lambda \big(\mathcal{L}_{A \to T} + \mathcal{L}_{T \to A}\big) + (1-\lambda) \big(\mathcal{L}_{A} + \mathcal{L}_{T} \big).
\end{equation}


This approach does not require any complex machinery to prevent collapse other than an EMA update of encoders from both modalities and the addition of the symmetry-breaking predictor.

\section{Experimental Setup}
\subsection{Datasets} \label{subsection: Datasets}

The list of datasets used in this work is reported in Table \ref{tab:datasets}.
We train SLAP on an internal private dataset of 260,000 pairs of full-length production-quality music tracks and professionally annotated captions (PrivateCaps \cite{manco2022contrastive}). Our retrieval testing datasets include two music-caption pair datasets. Specifically, we use MusicCaps \cite{agostinelli2023musiclm} and the Song Describer Dataset \cite{manco2023song} for text-music retrieval performance evaluation. MusicCaps contains 5,521 music clips, each accompanied by a detailed text description. The Song Describer Dataset includes 2-minute-long permissively licensed music clips with crowd-sourced single-sentence captions.

For zero-shot classification and downstream probing performance (See Sections \ref{subsection: Downstream probing}, \ref{subsection: Zero-shot performance}), we use the GTZAN dataset \cite{GTZAN} for music genre classification, which contains 1,000 audio tracks, each 30 seconds long, spanning 10 genres. We also use the MagnaTagATune (MTAT) dataset \cite{MTT}, an automatic tagging dataset consisting of 25,000 30-second music clips with 50 associated tags. Additionally, we employ the OpenMic dataset \cite{humphrey2018openmic} for fine-grained instrument tagging, containing annotations for instrument presence over 20,000 snippets.

\begin{table}
\centering
\resizebox{\columnwidth}{!}{
\begin{tabular}{llcc}
\toprule
\textbf{Task} & \textbf{Dataset} & \textbf{\# clips} & \textbf{Clip length} \\
\midrule
\multirow{1}{*}{Training} & PrivateCaps & 260k & full length \\
\midrule
\multirow{2}{*}{Retrieval} & MusicCaps~\cite{agostinelli2023musiclm} & 5.5k & 10 seconds \\
 & Song Describer~\cite{manco2023song} & 1k & 2 minutes \\
\midrule
\multirow{3}{*}{Probing/ZS} & GTZAN~\cite{GTZAN} & 1k & 30 seconds \\
 & MagnaTagATune~\cite{MTT} & 25k & 30 seconds \\
 & OpenMic~\cite{humphrey2018openmic} & 20k & 10 seconds \\
\bottomrule
\end{tabular}
}
\caption{Datasets used for training, retrieval, and downstream tasks (probing and zero-shot classification).}
\label{tab:datasets}
\end{table}

\input{tables/retrieval2}

\subsection{Model architecture}

We use the same architecture as in LAION-CLAP~\cite{wu2023large}, namely HTS-AT~\cite{chen2022hts} for the audio encoder and RoBERTa~\cite{liu2019roberta} for the text encoder.
This backbone configuration leads to an overall model size of 193M parameters.

The predictor architecture is a ReLU-activated multilayer perceptron (MLP) with batch normalization. By default, we employ one 4096-wide hidden layer for the MLP. Both the prediction and projection dimensions are 512, as in CLAP \cite{elizalde2023clap}. We leave the study of predictor architecture hyperparameters as well as the influence of encoder architecture to future work.

\subsection{Training details}

All models are trained on PrivateCaps using a batch size of 768 across 6 A100 GPUs with PyTorch automatic mixed precision, for 60 epochs.
We use linear warmup and cosine decay with a maximum learning rate of $4 \times 10^{-4}$ (scaled by batch size) and warmup over 1/10 of total epochs. We use SpecAugment as an audio augmentation \cite{park2019specaugment}, which improves performance slightly across tasks. We set EMA update rate $\tau = 0.95$, as we empirically find that lower values (compared to 0.996 in \cite{niizumi2021byol}) yield better retrieval performance, and leave full hyper-parameter tuning to future work. We experiment with both randomly initialized and pretrained checkpoints for HTS-AT\footnote{\href{https://github.com/LAION-AI/CLAP?tab=readme-ov-file}{\tt https://github.com/LAION-AI/CLAP}} and RoBERTa\footnote{\href{https://huggingface.co/FacebookAI/roberta-base}{\tt https://huggingface.co/FacebookAI/roberta-base}}, and initialize the HTS-AT backbone from a public checkpoint trained on AudioSet~\cite{audioset} when initializing from pretrained.

\section{Results}

\subsection{Multimodal retrieval}

We perform Audio to Text ($A \rightarrow T$) and Text to Audio ($T \rightarrow A$) retrieval across the test datasets described in Section \ref{subsection: Datasets}. We use predictions $q$ as key and query for SLAP models (see Section \ref{Subsection:Prediction vs projection retrieval}) and projections $z$ for CLAP models. We report Recall @(1,5,10), as well as Mean and Median Normalized Rank (MNR/MdNR) \cite{SampleMatch}. We report comparable results from relevant work in literature as well as our reproduced CLAP model.

We report retrieval results in Table \ref{tab:music-results}. We find that SLAP consistently outperforms both
\alain{MusCALL~\cite{manco2022contrastive}, which was also trained on PrivateCaps,}
and our own reproduced CLAP models. Notably, Recall and Normalized Rank are improved across all metrics except for the $T \rightarrow A$ task on MNR. Initializing from pretrained encoders leads to significantly better performance both for CLAP and SLAP.

\alain{
We also compare our results to LAION-CLAP~\cite{wu2023large}, which shares the same architecture. On Song Describer, all our models significantly outperform it, highlighting the importance of high-quality training data for music understanding.
On MusicCaps, whose audios are part of their training data, LAION-CLAP performs better; however, our best SLAP model, initialised from pretrained checkpoints, achieves close performances, particularly on $A\to T$.
}

\input{tables/projections}

\subsubsection{Prediction vs. projection retrieval}\label{Subsection:Prediction vs projection retrieval}

One variant of our approach is to use predictions instead of projections to perform retrieval. Given the loss formulation, both are possible since both exist in separate embedding spaces and the pretraining objective optimizes for four-way similarity. Here we observe the effect of using $q$ or $z$ as query and key for retrieval. Results are reported in Table \ref{Projection-Prediction}. We find that using projections instead of predictions leads to a small drop in performance, but both are viable for retrieval tasks, meaning there is no loss of information when compared to CLAP for projections.

\input{tables/probing}

\subsection{Downstream probing} \label{subsection: Downstream probing}

This section verifies the quality of the learned representations of SLAP by performing downstream probing on a range of tasks on representations \textit{before} the projection head which outputs $z$. For downstream task evaluation, we freeze the audio encoder and train shallow nonlinear probes (one ReLU-activated 512-wide hidden layer MLPs). Probes are trained with a learning rate of $10^{-4}$ with an early-stopping mechanism on validation loss.

Our results are reported in Table~\ref{tab:probingtable}. As above, we compare them to our CLAP baseline and MusCALL. We report the performance of  MATPAC~\cite{MATPAC}, the current self-supervised state-of-the-art on these tasks as well.
As a reference, we also indicate state-of-the-art performances for each task.


We observe that SLAP consistently outperforms CLAP and MusCALL, demonstrating the advantages of our target encoder/predictor strategy over contrastive learning.

On MTAT and OpenMic, SLAP surpasses MATPAC, which also employs a target encoder and predictor but is trained on audio alone. This suggests that language supervision is particularly beneficial for tagging tasks but is less useful for genre classification (82.9\% vs. 85.9\%).

On tagging tasks, SLAP achieves state-of-the-art results on MTAT (45.8\% vs. 41.1\%) and bridges the gap with best supervised methods on OpenMic (86.2\% vs. 86.7\%).



\subsection{Zero-shot performance} \label{subsection: Zero-shot performance}

Similar to shallow downstream probing, we evaluate the zero-shot performance of SLAP by retrieving the highest similarity prompts to audio using prompts as a class proxy on GTZAN, MTAT, and OpenMic. We report Zero-shot accuracy on GTZAN, as well as ROC-AUC and mAP for automatic tagging on MTAT, and mAR on OpenMic. As previous work has shown the importance of prompt engineering for zero-shot performance \cite{manco2022contrastive}, we evaluate across four prompts with which we wrap the target class ``\{\}'', ``\{\} music'', ``this sounds like \{\}'', and ``A \{\} track''.
We report the best of four scores across datasets for both CLAP, SLAP, and MusCALL~\cite{manco2022contrastive}.

We adapt another metric for zero-shot evaluation of multilabel approaches, which we call mean Average Recall (mAR). Previous approaches evaluate multi-tag zero-shot performance by using tag-wise cosine similarity as logits as input for standard metrics such as AUROC and mAP. These metrics are not robust to to cosine similarity distributions centered around 0.5 with low spreads for certain tasks, e.g. instrumentation.
mAR computes the mean rank of retrieved ground truth tags for a given audio:
\begin{equation}
    \mAR = \frac{1}{NK} \sum_{k=1}^K \sum_{n=1}^N R_n@k.
\end{equation}
where $K$ is the number of samples in the dataset, $N$ the number of tags, and $R_n@k$ recall at $k$ for tag $n$. We evaluate mAR on OpenMic. Results are reported in Table \ref{tab:zero} for CLAP, SLAP, and MusCALL \cite{manco2022contrastive}.

We observe that SLAP consistently outperforms or is on par with our CLAP baseline and MusCALL.
As for retrieval, a plausible cause is the better overlap between the audio and text spaces of our approach compared to contrastive ones. We measure this overlap in Section~\ref{modalitygap}.

\begin{table}[h]
\centering
\resizebox{\columnwidth}{!}{%
\begin{tabular}{lcccccc}
\toprule
 & \multicolumn{1}{c}{GTZAN} &  & \multicolumn{2}{c}{MTAT} &  & \multicolumn{1}{c}{OpenMIC} \\
\cmidrule{2-2} \cmidrule{4-5} \cmidrule{7-7}
Model & \multicolumn{1}{c}{Acc.} &  & \multicolumn{1}{c}{AUROC} & \multicolumn{1}{c}{mAP} &  &\multicolumn{1}{c}{mAR} \\
\midrule
SLAP   & \textbf{58.3} &  & \textbf{75.0} & \textbf{31.5} &  & \textbf{70.5} \\
CLAP   & 51.7 &  & \textbf{75.0} & 26.0 &  & 70.1 \\
\midrule
MusCALL \cite{manco2022contrastive} &
48.2 &  & 73.8 & 23.0 &  & 66.7 \\
\bottomrule
\end{tabular}%
}
\caption{Best-of-four zero-shot classification and tagging results for genre and instrument classification, and automatic tagging. Best scores are shown in \textbf{bold}.} 
\label{tab:zero}
\end{table}

\begin{figure}[t]
    \centering
    \includegraphics[width=\linewidth]{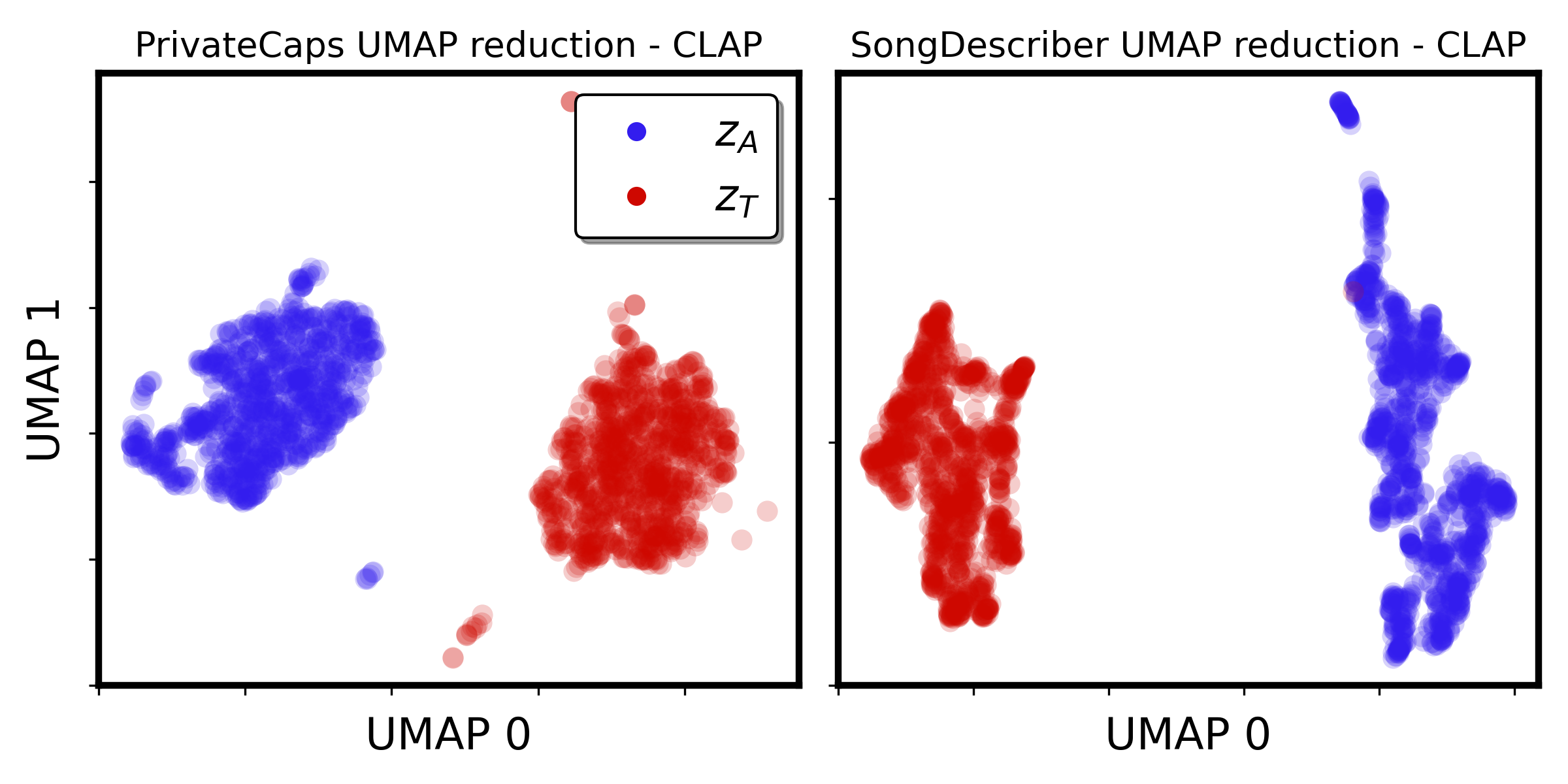}
    \includegraphics[width=\linewidth]{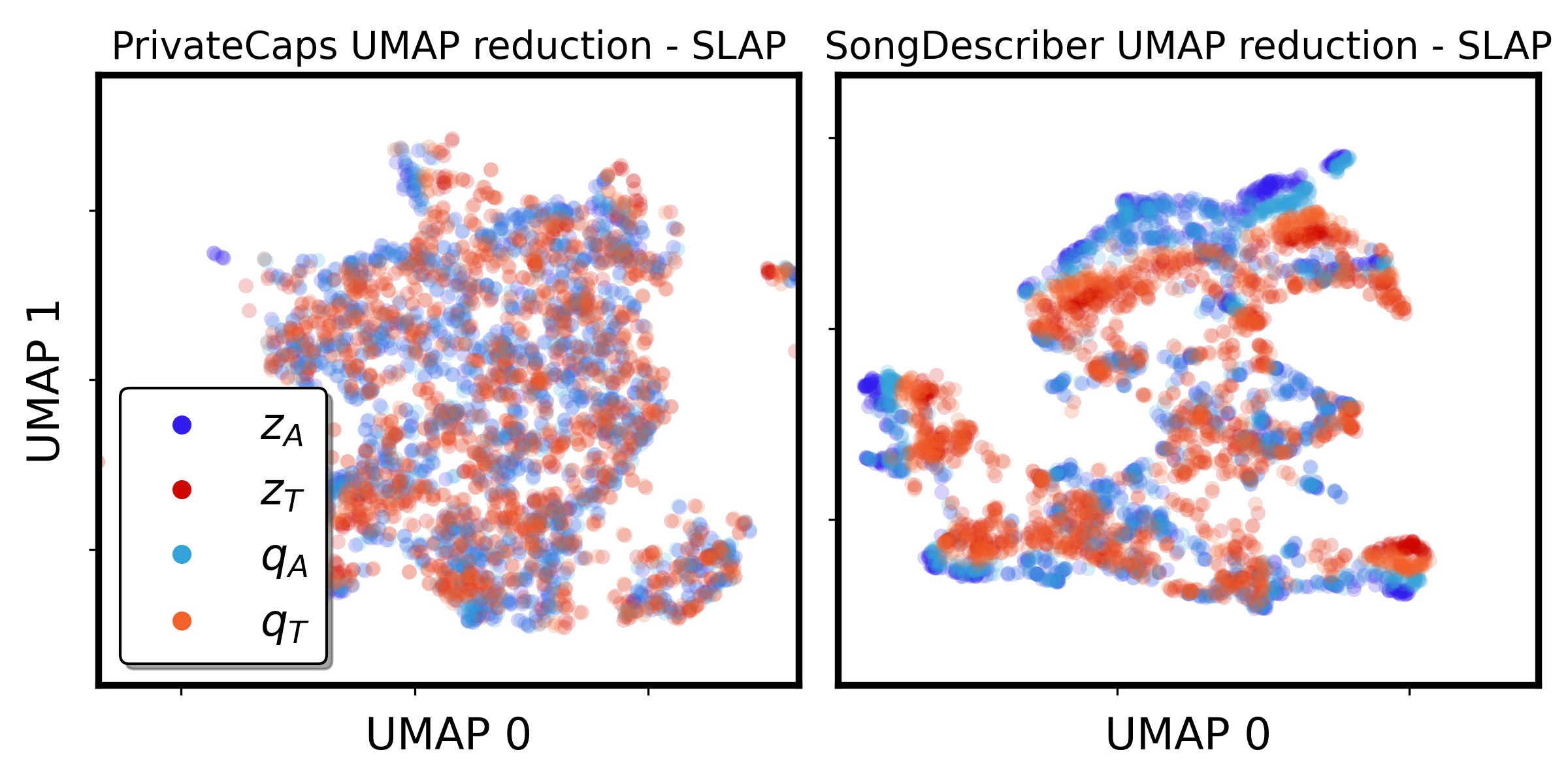}
    \caption{
        UMAP reduction of the CLAP (\textbf{top}) and SLAP (\textbf{bottom}) embeddings obtained from PrivateCaps (\textbf{left}) and Song Describer (\textbf{right}).
    }
    \label{fig:UMAPreductions}
\end{figure}


\begin{figure}
    \includegraphics[width=0.95\columnwidth]{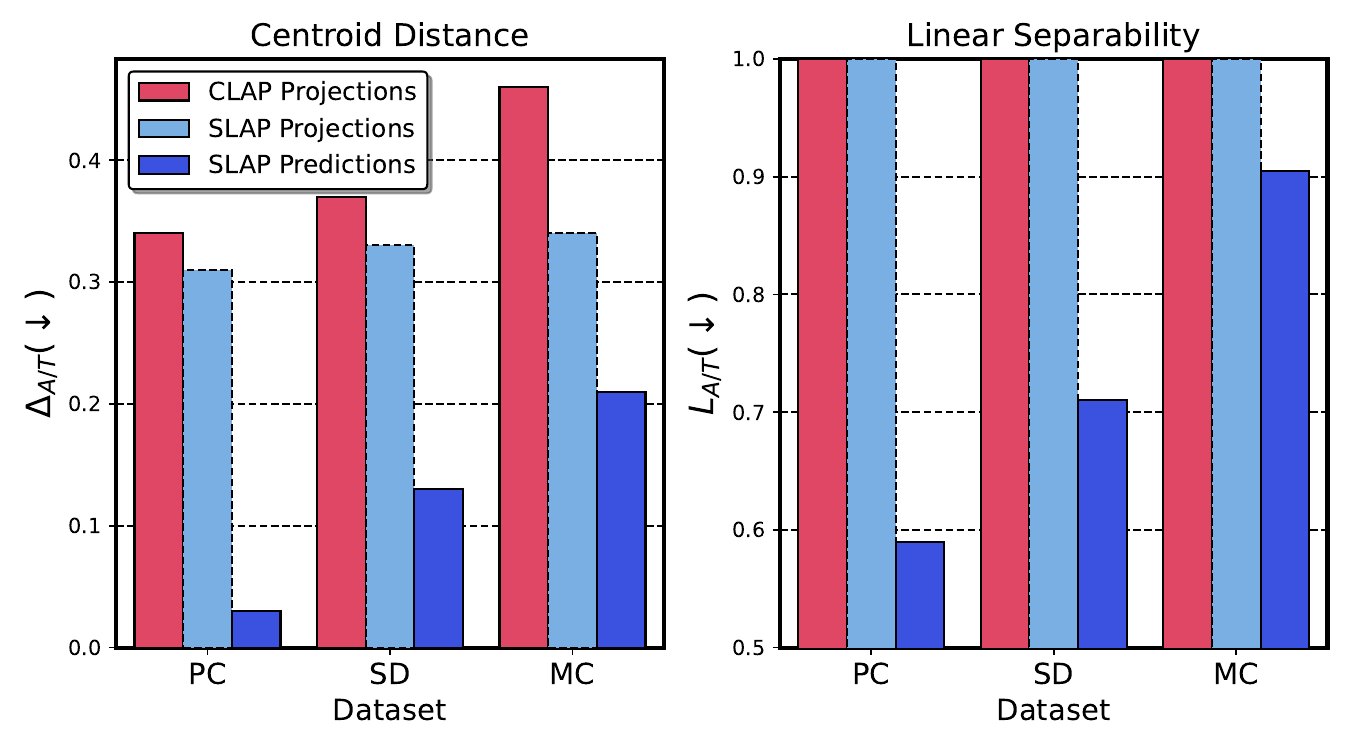}
    \caption{Centroid Distance (\textbf{left}) and Linear separability (\textbf{right}) for SLAP and CLAP on PrivateCaps (PC), Song Describer (SD) and MusicCaps (MC). Lower is better.}
    \label{fig:modalitygap}
\end{figure}

\subsection{Modality gap}\label{modalitygap}

As discussed in Section \ref{sec:limits}, one notable issue with MCL is the presence of a modality gap~\cite{liang2022mind,Fahim2024}.
While it is unclear how detrimental the modality gap is to performance for retrieval, it has been shown that ``closing the gap'' is beneficial at least for generative applications \cite{nistal2024improving}.
Intuitively, having disjoint manifolds of audio and text representations allows their usage for cosine-similarity Max Inner Product Search, but is not beneficial for a shared joint understanding of text and audio. Multimodal contrastive models are subject to a modality gap, likely attributable to the formulation of contrastive loss \cite{Fahim2024}.
Not being contrastive, we posit that our method should be less prone to creating such a gap.
Figure \ref{fig:UMAPreductions} shows the UMAP reduction of projections $z_T,z_A$ and $q_T,q_A$ from PrivateCaps and Song Describer, for both CLAP and SLAP.
Although the modality gap is immediately apparent for CLAP, it seems much less pronounced for SLAP.

As in \cite{Fahim2024} and \cite{liang2022mind}, we measure this by computing the linear separability $L_{A/T}$ between the respective manifolds of audio and text in the latent space.
For both PrivateCaps, MusicCaps and Song Describer, we report in Figure \ref{fig:modalitygap} the training accuracy of a Logistic Regression classifier trained to distinguish an audio embedding from a text embedding.
We also measure the Euclidean distance $\Delta_{A/T}$ between the centroids of each modality.

These observations confirm the modality gap is larger, both in terms of distance and linear separability, for CLAP than for SLAP. More so, it is nearly nonexistent for SLAP on both PC and SD. Although the gap is larger for MC than for PC and SD, this can be attributed to a domain shift between datasets, likely attributable to the noisiness and longer captions of MC.

\subsection{Scalability properties}

\begin{figure}
    \centering
    \includegraphics[width=.73\columnwidth]{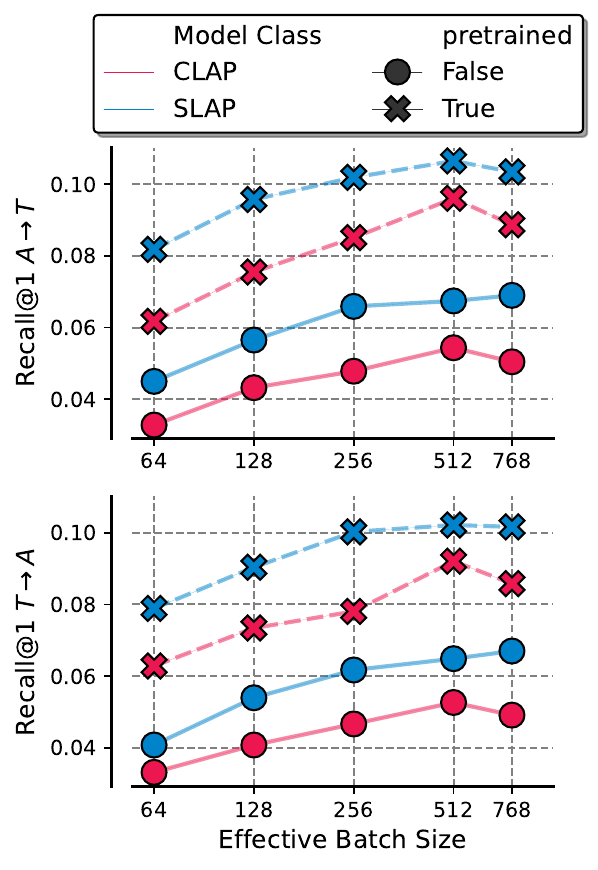}
    \caption{Scaling batch size for CLAP and SLAP.}
    \label{fig:Batch size scaling}
\end{figure}

We compare the scalability of SLAP and CLAP, focusing on batch size. A key limitation of contrastive models like CLAP is that their loss depends on all negatives in a batch, making gradient accumulation ineffective—since the gradient does not scale linearly with batch size. In contrast, SLAP requires no negatives and can theoretically be scaled to vastly large batch sizes, notably due to the lesser memory footprint and the availability of gradient accumulation. We study the effect of batch size on retrieval performance by reporting $A\rightarrow T$ and $T\rightarrow A$ R@1 on PrivateCaps for batch sizes from 64 to 768. For CLAP, batch size is scaled directly. For SLAP, we fix a base batch of 128 and accumulate gradients over $N = B // 128$ steps to reach target batch size $B$. We empirically confirm that gradient accumulation in SLAP is equivalent to true batch scaling. Results are shown in Figure~\ref{fig:Batch size scaling}.


\subsection{Role of Intermodal and Intramodal Loss terms}\label{subsection: role of intermodal and intramodal loss terms}

\begin{figure}[h]
    \centering
    \includegraphics[width=1\columnwidth]{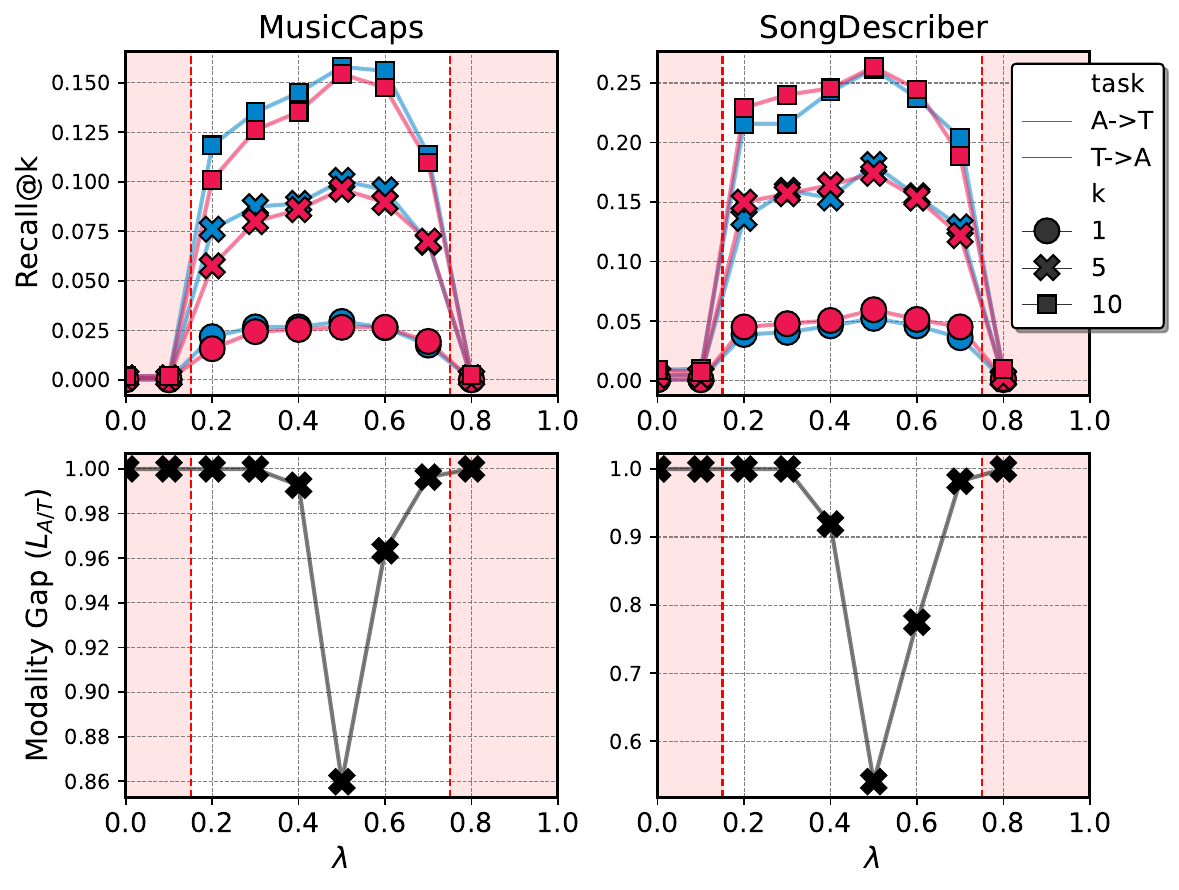}
    \caption{Tuning of intramodality loss balancing weight $\lambda$ for the SLAP objective, measuring both $T\rightarrow A$ and $A\rightarrow T$ retrieval performance and $L_{A/T}$.}
    \label{fig:intermodality loss}
\end{figure}

One key component of our approach is the multiple loss terms between the different EMA encoders of the training setup. We find that empirically, the absence of these losses $\mathcal{L}_A$ and $\mathcal{L}_T$ leads to collapse of the model.
We show this in Figure \ref{fig:intermodality loss} by reporting recall@$k$ and linear separability between modalities
for different balancing weights $\lambda$.

We find that within a range of $\lambda$, SLAP is robust for retrieval. Values of $\lambda$ outside [0.2, 0.7] lead to almost systematic collapse due to the lack of meaningful constraints between or within modalities. Despite good retrieval robustness, we find that any $\lambda \neq 0.5$ yields a rapidly increasing modality gap. Only a balanced contribution of intra and inter-modality loss terms leads to the significant reduction of the modality gap observed in Section \ref{modalitygap}.

\section{Conclusion and future work}


We propose a novel approach to training multimodal joint embedding spaces to align music and text representation: Siamese Language Audio Pretraining (SLAP). SLAP outperforms contrastive models on tasks including text-music retrieval, downstream probing, and zero-shot music understanding. It offers key advantages over contrastive approaches, including negative-free training for better scalability and substantial reduction of the modality gap. Without relying on extensive data augmentation, architectural tuning, or hyperparameter optimization, SLAP matches state-of-the-art performance—demonstrating its potential as a scalable alternative to CLAP-style training.

\alain{CLIP/CLAP being a core component of many text-conditioned models, several works have proposed to improve them,} 
e.g., by exploring alternative architectures~\cite{elizalde2023clap,CyCLAP}, scaling-up training data~\cite{ghosh2025reclap,AudioFlamingo2}, promoting semantic or linguistic invariance~\cite{MaskCLIP,AudioFlamingo2} or improving the sensitivity to fine-grained temporal events~\cite{yuan2024t,wu2024collap}.
\alain{Importantly, these improvements typically do not rely on the contrastive loss,}
and SLAP could probably benefit from them as well.

Finally, while we demonstrate the effectiveness of our method on music-related tasks, it is generalizable and could be applied to other modalities such as images or general audio.
We therefore believe that this approach could lead to many applications beyond the scope of our paper.

\section{Acknowledgement}

This work is supported by the EPSRC UKRI Centre for
Doctoral Training in Artificial Intelligence and Music
(EP/S022694/1) and Universal Music Group.

\bibliography{refs}








\end{document}

%% file: tables/retrieval2.tex
\begin{table*}[t]
\resizebox{\textwidth}{!}{%
\begin{tabular}{lccccccccccccccccccccc}
\toprule
                           &            & \multicolumn{10}{c|}{Song Describer}                                                                                        & \multicolumn{10}{c}{MusicCaps}                                                                                            \\ \cline{3-22} 
                           &            & \multicolumn{5}{c|}{$A \rightarrow T$}                       & \multicolumn{5}{c|}{$T \rightarrow A$}                       & \multicolumn{5}{c|}{$A \rightarrow T$}                       & \multicolumn{5}{c}{$T \rightarrow A$}                      \\ \cmidrule{3-22} 
                           &            & \multicolumn{3}{c}{Recall $(\uparrow)$} & \multicolumn{2}{c|}{Norm. rank ($\downarrow$)}  & \multicolumn{3}{c}{Recall $(\uparrow)$} & \multicolumn{2}{c|}{Norm. rank ($\downarrow$)}  & \multicolumn{3}{c}{Recall $(\uparrow)$} & \multicolumn{2}{c|}{Norm. rank ($\downarrow$)}  & \multicolumn{3}{c}{Recall $(\uparrow)$} & \multicolumn{2}{c}{Norm. rank ($\downarrow$)} \\ \cmidrule{3-22} 
Model                      & Pre. & 1       & 5      & 10      & Med. & \multicolumn{1}{c|}{Mean} & 1       & 5      & 10      & Med. & \multicolumn{1}{c|}{Mean} & 1       & 5      & 10      & Med. & \multicolumn{1}{c|}{Mean} & 1       & 5      & 10      & Med.           & Mean          \\
\midrule

\multirow{1}{*}{SLAP} & \redcross & \textbf{3.3} & \textbf{13.0} & \textbf{19.7} & \textbf{5.3} &   \multicolumn{1}{c|}{\textbf{12.3}} & \textbf{3.7} & \textbf{12.5} & \textbf{19.4} & \textbf{5.4} & \multicolumn{1}{c|}{\textbf{13.8}} & \textbf{1.6} & \textbf{5.3} & \textbf{7.5} & \textbf{4.8} & \multicolumn{1}{c|}{\textbf{14.5}} & \textbf{1.2} & \textbf{4.9} & \textbf{7.9} &  \textbf{4.4} &  \textbf{13.3} \\

CLAP & \redcross & 3.1 & 9.6 & 14.9 & 7.3 & \multicolumn{1}{c|}{14.2}  & 2.7 & 9.7 & 15.8 & 6.7 & \multicolumn{1}{c|}{16.4} & 0.9 & 3.4 & 5.2 & 8.8 & \multicolumn{1}{c|}{18.2}  & 0.7 & 3.2 & 5.2 & 7.5 & 17.4 \\

SLAP &        \greencheck    & \textbf{5.7}  & \textbf{18.1} & \textbf{26.6} &\textbf{ 3.2 }& \multicolumn{1}{c|}{\textbf{8.9}} & \textbf{6.0} & \textbf{18.1}       & \textbf{ 26.4}       &   \textbf{3.6}  & \multicolumn{1}{c|}{10.5}     & \textbf{3.1} & \textbf{10.1} & \textbf{15.4} &  \textbf{1.9}   & \multicolumn{1}{c|}{\textbf{7.7}}     &      \textbf{3.0}   & \textbf{9.6} & \textbf{15.4}  & \textbf{1.9}  & 7.7 \\

CLAP &     \greencheck       & 5.3 & 14.9  & 22.2 &  4.3   & \multicolumn{1}{c|}{9.2} & 5.7 & 16.8 & 24.1 &  4.3   & \multicolumn{1}{c|}{\textbf{9.7}} & 2.8 & 8.3 &     10.4 &  3.0   & \multicolumn{1}{c|}{10.0}     & 2.8  & 8.7 & 14.0 & 2.2 & \textbf{7.6} \\
\midrule
\multicolumn{2}{l}{MusCALL} &
1.4 & 7.0 & 14.4 & 7.8 &  \multicolumn{1}{c|}{14.7} & 1.9 & 6.8 & 12.1 & 8.8 & \multicolumn{1}{c|}{19.9} &  1.5     &  7.1   &  10.8  & 9.3  & \multicolumn{1}{c|}{17.3 }     &  0.4   &  1.6  &  3.0  & 12.8 & 21.1 \\
\multicolumn{2}{l}{LAION-CLAP$^\dagger$}            &   1.9   &   7.4   &   11.4 &   9.4  & \multicolumn{1}{c|}{18.7}     &   1.6      &  5.3    &  9.5       &   9.2  & \multicolumn{1}{c|}{18.2}     &   3.2      &      10.4  &   16.0      &  2.0   & \multicolumn{1}{c|}{6.9}     &   3.7      & 11.1        &    17.0     &    1.6           &    5.5           \\

\bottomrule
\end{tabular}%
}
\begin{flushleft}
    \vspace{-2mm}
    \footnotesize$^\dagger$ \alain{Note that LAION-CLAP is trained on a large dataset that includes AudioSet~\cite{audioset}, from which the audio samples of MusicCaps were extracted.}
    \vspace{-5mm}
\end{flushleft}
\caption{
    Text-Music retrieval results. \textbf{Bolded} results indicate best results between CLAP and SLAP for either models initialized from pretrained checkpoints (\greencheck) or from scratch (\redcross). All metrics are in percentages.
}
\label{tab:music-results}
\end{table*}

%% file: tables/projections.tex
\begin{table}[h]
    \centering
    \resizebox{\columnwidth}{!}{%
    \begin{tabular}{lcccc|cccc}
    \toprule
    & \multicolumn{4}{c}{Song Describer} & \multicolumn{4}{c}{MusicCaps} \\
    \cmidrule{2-9} 
     & \multicolumn{2}{c}{$A \rightarrow T$} & \multicolumn{2}{c}{$T \rightarrow A$} & \multicolumn{2}{c}{$A \rightarrow T$} & \multicolumn{2}{c}{$T \rightarrow A$} \\
    \midrule
    Anchor & R@5 & MdNR & R@5 & MdNR & R@5 & MdNR & R@5 & MdNR \\ \midrule
    $z$ & 17.8 & 3.6 & 17.0 & 3.8 & 9.6 & 2.1 & 9.4 & 2.3 \\
    $q$ & \textbf{18.1} & \textbf{3.2} & \textbf{18.1} & \textbf{3.6} & \textbf{10.1} & \textbf{1.9} & \textbf{9.6} & \textbf{1.9} \\
    \bottomrule
    \end{tabular}%
    }
    \caption{Retrieval results using \alain{queries} $q$ or projection embeddings $z$ as anchors. Best results are in \textbf{bold}.}
    \label{Projection-Prediction}
\end{table}

%% file: tables/probing.tex
\begin{table}[h]
    \centering
    \resizebox{\columnwidth}{!}{%
    \begin{tabular}{lcccccc}
    \toprule
     & \multicolumn{1}{c}{GTZAN} &  & \multicolumn{2}{c}{MTAT} &  & \multicolumn{1}{c}{OpenMic} \\
    \cmidrule{2-2} \cmidrule{4-5} \cmidrule{7-7} 
    Model     & \multicolumn{1}{c}{Acc.} &  & \multicolumn{1}{c}{AUROC} & \multicolumn{1}{c}{mAP} &  & \multicolumn{1}{c}{mAP}  \\
    \midrule
    SLAP &
    82.9 &  & \underline{92.0} & \textbf{45.8} &  & \underline{86.2} \\
    CLAP &
    80.9  &  & 91.7 & \underline{45.1} &  & 85.8 \\
    \midrule
    MusCALL & 74.5 &  & 91.5 & 44.4 &  & 79.7   \\
    \midrule
    MATPAC &
    \underline{85.9} &  & 91.6 & 41.1 &  & 85.4 \\
    SOTA &
    \textbf{87.4}~\cite{koutini2022efficient} &  & \textbf{92.7}~\cite{mulan}   & 41.4~\cite{castellon2021codified} &  & \textbf{86.7}~\cite{chen2022beats} \\ 
    \bottomrule
    \end{tabular}%
    }
    \caption{
        Probing results on frozen representations for genre classification (GTZAN), automatic tagging (MTAT), and Instrument tagging (OpenMic).
        Best scores are \textbf{bolded} and second-to-best are \underline{underlined}.
    }
    \label{tab:probingtable}
\end{table}

%% file: main.bbl
\begin{thebibliography}{10}
\providecommand{\url}[1]{#1}
\csname url@samestyle\endcsname
\providecommand{\newblock}{\relax}
\providecommand{\bibinfo}[2]{#2}
\providecommand{\BIBentrySTDinterwordspacing}{\spaceskip=0pt\relax}
\providecommand{\BIBentryALTinterwordstretchfactor}{4}
\providecommand{\BIBentryALTinterwordspacing}{\spaceskip=\fontdimen2\font plus
\BIBentryALTinterwordstretchfactor\fontdimen3\font minus \fontdimen4\font\relax}
\providecommand{\BIBforeignlanguage}[2]{{%
\expandafter\ifx\csname l@#1\endcsname\relax
\typeout{** WARNING: IEEEtran.bst: No hyphenation pattern has been}%
\typeout{** loaded for the language `#1'. Using the pattern for}%
\typeout{** the default language instead.}%
\else
\language=\csname l@#1\endcsname
\fi
#2}}
\providecommand{\BIBdecl}{\relax}
\BIBdecl

\bibitem{clip}
A.~Radford, J.~W. Kim, C.~Hallacy \emph{et~al.}, ``Learning transferable visual models from natural language supervision,'' in \emph{International conference on machine learning}.\hskip 1em plus 0.5em minus 0.4em\relax PMLR, 2021, pp. 8748--8763.

\bibitem{elizalde2023clap}
B.~Elizalde, S.~Deshmukh, M.~Al~Ismail \emph{et~al.}, ``Clap learning audio concepts from natural language supervision,'' in \emph{ICASSP 2023-2023 IEEE International Conference on Acoustics, Speech and Signal Processing (ICASSP)}.\hskip 1em plus 0.5em minus 0.4em\relax IEEE, 2023, pp. 1--5.

\bibitem{wu2024collap}
J.~Wu, W.~Li, Z.~Novack \emph{et~al.}, ``Collap: Contrastive long-form language-audio pretraining with musical temporal structure augmentation,'' in \emph{ICASSP 2025-2025 IEEE International Conference on Acoustics, Speech and Signal Processing (ICASSP)}.\hskip 1em plus 0.5em minus 0.4em\relax IEEE, 2025, pp. 1--5.

\bibitem{komatsu2025aligned}
T.~Komatsu, H.~Munakata, T.~Hasumi \emph{et~al.}, ``Aligned contrastive learning for text-to-music retrieval,'' in \emph{ICASSP 2025-2025 IEEE International Conference on Acoustics, Speech and Signal Processing (ICASSP)}.\hskip 1em plus 0.5em minus 0.4em\relax IEEE, 2025, pp. 1--5.

\bibitem{yuan2024t}
Y.~Yuan, Z.~Chen, X.~Liu \emph{et~al.}, ``T-clap: Temporal-enhanced contrastive language-audio pretraining,'' in \emph{2024 IEEE 34th International Workshop on Machine Learning for Signal Processing (MLSP)}.\hskip 1em plus 0.5em minus 0.4em\relax IEEE, 2024, pp. 1--6.

\bibitem{manco2024augment}
I.~Manco, J.~Salamon, and O.~Nieto, ``Augment, drop \& swap: Improving diversity in llm captions for efficient music-text representation learning,'' in \emph{Proceedings of the 25th International Society for Music Information Retrieval Conference (ISMIR)}, 2024.

\bibitem{Fahim2024}
\BIBentryALTinterwordspacing
A.~Fahim, A.~Murphy, and A.~Fyshe, ``{Its Not a Modality Gap: Characterizing and Addressing the Contrastive Gap},'' may 2024. [Online]. Available: \url{http://arxiv.org/abs/2405.18570}
\BIBentrySTDinterwordspacing

\bibitem{liang2022mind}
V.~W. Liang, Y.~Zhang, Y.~Kwon \emph{et~al.}, ``Mind the gap: Understanding the modality gap in multi-modal contrastive representation learning,'' \emph{Advances in Neural Information Processing Systems}, vol.~35, pp. 17\,612--17\,625, 2022.

\bibitem{pham2023combined}
H.~Pham, Z.~Dai, G.~Ghiasi \emph{et~al.}, ``Combined scaling for zero-shot transfer learning,'' \emph{Neurocomputing}, vol. 555, p. 126658, 2023.

\bibitem{grill2020bootstrap}
J.-B. Grill, F.~Strub, F.~Altch{\'e} \emph{et~al.}, ``Bootstrap your own latent-a new approach to self-supervised learning,'' \emph{Advances in neural information processing systems}, vol.~33, pp. 21\,271--21\,284, 2020.

\bibitem{niizumi2021byol}
D.~Niizumi, D.~Takeuchi, Y.~Ohishi \emph{et~al.}, ``Byol for audio: Self-supervised learning for general-purpose audio representation,'' in \emph{2021 International Joint Conference on Neural Networks (IJCNN)}.\hskip 1em plus 0.5em minus 0.4em\relax IEEE, 2021, pp. 1--8.

\bibitem{chen2020simple}
T.~Chen, S.~Kornblith, M.~Norouzi \emph{et~al.}, ``A simple framework for contrastive learning of visual representations,'' in \emph{International conference on machine learning}.\hskip 1em plus 0.5em minus 0.4em\relax PMLR, 2020, pp. 1597--1607.

\bibitem{spijkervet2021contrastive}
J.~Spijkervet and J.~A. Burgoyne, ``Contrastive learning of musical representations,'' in \emph{Proceedings of the 22nd International Society for Music Information Retrieval Conference (ISMIR)}, 2021.

\bibitem{VATT}
\BIBentryALTinterwordspacing
H.~Akbari, L.~Yuan, R.~Qian, W.~H. Chuang, S.~F. Chang, Y.~Cui, and B.~Gong, ``{VATT: Transformers for Multimodal Self-Supervised Learning from Raw Video, Audio and Text},'' in \emph{Advances in Neural Information Processing Systems}, vol.~29.\hskip 1em plus 0.5em minus 0.4em\relax Neural information processing systems foundation, apr 2021, pp. 24\,206--24\,221. [Online]. Available: \url{https://arxiv.org/abs/2104.11178v3}
\BIBentrySTDinterwordspacing

\bibitem{L3}
\BIBentryALTinterwordspacing
R.~Arandjelovic and A.~Zisserman, ``{Look, Listen and Learn},'' in \emph{Proceedings of the IEEE International Conference on Computer Vision}, vol. 2017-Octob.\hskip 1em plus 0.5em minus 0.4em\relax Institute of Electrical and Electronics Engineers Inc., may 2017, pp. 609--617. [Online]. Available: \url{https://arxiv.org/abs/1705.08168v2}
\BIBentrySTDinterwordspacing

\bibitem{manco2022contrastive}
I.~Manco, E.~Benetos, E.~Quinton \emph{et~al.}, ``Contrastive audio-language learning for music,'' in \emph{Proceedings of the 23rd International Society for Music Information Retrieval Conference, {ISMIR} 2022, Bengaluru, India, December 4-8, 2022}, P.~Rao, H.~A. Murthy, A.~Srinivasamurthy, R.~M. Bittner, R.~C. Repetto, M.~Goto, X.~Serra, and M.~Miron, Eds., 2022, pp. 640--649.

\bibitem{mulan}
Q.~Huang, A.~Jansen, J.~Lee \emph{et~al.}, ``Mulan: {A} joint embedding of music audio and natural language,'' in \emph{Proceedings of the 23rd International Society for Music Information Retrieval Conference, {ISMIR} 2022, Bengaluru, India, December 4-8, 2022}, P.~Rao, H.~A. Murthy, A.~Srinivasamurthy, R.~M. Bittner, R.~C. Repetto, M.~Goto, X.~Serra, and M.~Miron, Eds., 2022, pp. 559--566.

\bibitem{zhu2024cacophony}
G.~Zhu, J.~Darefsky, and Z.~Duan, ``Cacophony: An improved contrastive audio-text model,'' \emph{IEEE/ACM Transactions on Audio, Speech, and Language Processing}, 2024.

\bibitem{guzhov2022audioclip}
A.~Guzhov, F.~Raue, J.~Hees \emph{et~al.}, ``Audioclip: Extending clip to image, text and audio,'' in \emph{ICASSP 2022-2022 IEEE International Conference on Acoustics, Speech and Signal Processing (ICASSP)}.\hskip 1em plus 0.5em minus 0.4em\relax IEEE, 2022, pp. 976--980.

\bibitem{girdhar2023imagebind}
R.~Girdhar, A.~El-Nouby, Z.~Liu \emph{et~al.}, ``Imagebind: One embedding space to bind them all,'' in \emph{Proceedings of the IEEE/CVF conference on computer vision and pattern recognition}, 2023, pp. 15\,180--15\,190.

\bibitem{GRAM}
\BIBentryALTinterwordspacing
G.~Cicchetti, E.~Grassucci, L.~Sigillo, and D.~Comminiello, ``{Gramian Multimodal Representation Learning and Alignment},'' dec 2024. [Online]. Available: \url{http://arxiv.org/abs/2412.11959}
\BIBentrySTDinterwordspacing

\bibitem{ramesh2022hierarchical}
A.~Ramesh, P.~Dhariwal, A.~Nichol \emph{et~al.}, ``Hierarchical text-conditional image generation with clip latents,'' \emph{arXiv preprint arXiv:2204.06125}, vol.~1, no.~2, p.~3, 2022.

\bibitem{alayrac2022flamingo}
J.-B. Alayrac, J.~Donahue, P.~Luc \emph{et~al.}, ``Flamingo: a visual language model for few-shot learning,'' \emph{Advances in neural information processing systems}, vol.~35, pp. 23\,716--23\,736, 2022.

\bibitem{AudioFlamingo}
\BIBentryALTinterwordspacing
Z.~Kong, A.~Goel, R.~Badlani \emph{et~al.}, ``{Audio Flamingo: A Novel Audio Language Model with Few-Shot Learning and Dialogue Abilities},'' in \emph{Proceedings of Machine Learning Research}, vol. 235.\hskip 1em plus 0.5em minus 0.4em\relax ML Research Press, feb 2024, pp. 25\,125--25\,148. [Online]. Available: \url{https://arxiv.org/abs/2402.01831v3}
\BIBentrySTDinterwordspacing

\bibitem{ghosh2025reclap}
S.~Ghosh, S.~Kumar, C.~K.~R. Evuru \emph{et~al.}, ``Reclap: Improving zero shot audio classification by describing sounds,'' in \emph{ICASSP 2025-2025 IEEE International Conference on Acoustics, Speech and Signal Processing (ICASSP)}.\hskip 1em plus 0.5em minus 0.4em\relax IEEE, 2025, pp. 1--5.

\bibitem{li2025drcap}
X.~Li, W.~Chen, Z.~Ma \emph{et~al.}, ``Drcap: Decoding clap latents with retrieval-augmented generation for zero-shot audio captioning,'' in \emph{ICASSP 2025-2025 IEEE International Conference on Acoustics, Speech and Signal Processing (ICASSP)}.\hskip 1em plus 0.5em minus 0.4em\relax IEEE, 2025, pp. 1--5.

\bibitem{ghosh2024recap}
S.~Ghosh, S.~Kumar, C.~K.~R. Evuru \emph{et~al.}, ``Recap: Retrieval-augmented audio captioning,'' in \emph{ICASSP 2024-2024 IEEE International Conference on Acoustics, Speech and Signal Processing (ICASSP)}.\hskip 1em plus 0.5em minus 0.4em\relax IEEE, 2024, pp. 1161--1165.

\bibitem{evans2024fast}
Z.~Evans, C.~Carr, J.~Taylor \emph{et~al.}, ``Fast timing-conditioned latent audio diffusion,'' in \emph{Proceedings of the 41st International Conference on Machine Learning}, 2024, pp. 12\,652--12\,665.

\bibitem{evans2024long}
Z.~Evans, J.~D. Parker, C.~Carr \emph{et~al.}, ``Long-form music generation with latent diffusion,'' in \emph{Proceedings of the 25th International Society for Music Information Retrieval Conference (ISMIR)}, 2024.

\bibitem{diffariff}
J.~Nistal, M.~Pasini, C.~Aouameur \emph{et~al.}, ``Diff-a-riff: Musical accompaniment co-creation via latent diffusion models,'' in \emph{ISMIR, 2024}, 2024.

\bibitem{liu2023audioldm}
H.~Liu, Z.~Chen, Y.~Yuan \emph{et~al.}, ``{AudioLDM}: Text-to-audio generation with latent diffusion models,'' in \emph{Proc. ICML}, 2023.

\bibitem{agostinelli2023musiclm}
A.~Agostinelli, T.~I. Denk, Z.~Borsos \emph{et~al.}, ``{MusicLM}: Generating music from text,'' \emph{arXiv preprint arXiv:2301.11325}, 2023.

\bibitem{SigLIP}
\BIBentryALTinterwordspacing
X.~Zhai, B.~Mustafa, A.~Kolesnikov \emph{et~al.}, ``{Sigmoid Loss for Language Image Pre-Training},'' in \emph{Proceedings of the IEEE International Conference on Computer Vision}.\hskip 1em plus 0.5em minus 0.4em\relax Institute of Electrical and Electronics Engineers Inc., mar 2023, pp. 11\,941--11\,952. [Online]. Available: \url{https://arxiv.org/abs/2303.15343v4}
\BIBentrySTDinterwordspacing

\bibitem{Assran2023}
\BIBentryALTinterwordspacing
M.~Assran, R.~Balestriero, Q.~Duval \emph{et~al.}, ``{The Hidden Uniform Cluster Prior in Self-Supervised Learning},'' in \emph{11th International Conference on Learning Representations, ICLR 2023}, oct 2023. [Online]. Available: \url{http://arxiv.org/abs/2210.07277}
\BIBentrySTDinterwordspacing

\bibitem{Shi2023}
P.~Y. Shi, M.~Welle, M.~Bj{\"{o}}rkman \emph{et~al.}, ``{Understanding the Modality Gap in CLIP},'' in \emph{International Conference on Learning Reprsentations}, no. 2023, 2023.

\bibitem{chen2021exploring}
X.~Chen and K.~He, ``Exploring simple siamese representation learning,'' in \emph{Proceedings of the IEEE/CVF conference on computer vision and pattern recognition}, 2021, pp. 15\,750--15\,758.

\bibitem{assranSelfSupervisedLearningImages2023}
M.~Assran, Q.~Duval, I.~Misra \emph{et~al.}, ``Self-{Supervised} {Learning} from {Images} with a {Joint}-{Embedding} {Predictive} {Architecture},'' in \emph{2023 {IEEE}/{CVF} {Conference} on {Computer} {Vision} and {Pattern} {Recognition} ({CVPR})}, jun 2023, pp. 15\,619--15\,629, iSSN: 2575-7075.

\bibitem{Tian2021}
\BIBentryALTinterwordspacing
Y.~Tian, X.~Chen, and S.~Ganguli, ``{Understanding Self-Supervised Learning Dynamics without Contrastive Pairs},'' \emph{Proceedings of Machine Learning Research}, vol. 139, pp. 10\,268--10\,278, feb 2021. [Online]. Available: \url{http://arxiv.org/abs/2102.06810}
\BIBentrySTDinterwordspacing

\bibitem{DINOv2}
\BIBentryALTinterwordspacing
M.~Oquab, T.~Darcet, T.~Moutakanni \emph{et~al.}, ``{DINOv2: Learning Robust Visual Features without Supervision},'' apr 2023. [Online]. Available: \url{http://arxiv.org/abs/2304.07193}
\BIBentrySTDinterwordspacing

\bibitem{baevskiData2vecGeneralFramework2022}
A.~Baevski, W.-N. Hsu, Q.~Xu \emph{et~al.}, ``\BIBforeignlanguage{en}{data2vec: {A} {General} {Framework} for {Self}-supervised {Learning} in {Speech}, {Vision} and {Language}},'' in \emph{\BIBforeignlanguage{en}{Proceedings of {Machine} {Learning} {Research}}}, Baltimore, MD, USA, 2022.

\bibitem{ATST}
\BIBentryALTinterwordspacing
X.~Li and X.~Li, ``{ATST: Audio Representation Learning with Teacher-Student Transformer},'' in \emph{Proceedings of the Annual Conference of the International Speech Communication Association, INTERSPEECH}, vol. 2022-Septe.\hskip 1em plus 0.5em minus 0.4em\relax International Speech Communication Association, apr 2022, pp. 4172--4176. [Online]. Available: \url{http://dx.doi.org/10.21437/Interspeech.2022-10126}
\BIBentrySTDinterwordspacing

\bibitem{niizumiMaskedModelingDuo2023}
D.~Niizumi, D.~Takeuchi, Y.~Ohishi \emph{et~al.}, ``\BIBforeignlanguage{en}{Masked {Modeling} {Duo}: {Learning} {Representations} by {Encouraging} {Both} {Networks} to {Model} the {Input}},'' in \emph{\BIBforeignlanguage{en}{{ICASSP} 2023 - 2023 {IEEE} {International} {Conference} on {Acoustics}, {Speech} and {Signal} {Processing} ({ICASSP})}}.\hskip 1em plus 0.5em minus 0.4em\relax Rhodes Island, Greece: IEEE, jun 2023, pp. 1--5.

\bibitem{MATPAC}
A.~Quelennec, P.~Chouteau, G.~Peeters \emph{et~al.}, ``Masked latent prediction and classification for self-supervised audio representation learning,'' in \emph{ICASSP 2025 - 2025 IEEE International Conference on Acoustics, Speech and Signal Processing (ICASSP)}, 2025, pp. 1--5.

\bibitem{manco2023song}
I.~Manco, B.~Weck, S.~Doh \emph{et~al.}, ``The song describer dataset: a corpus of audio captions for music-and-language evaluation,'' NeurIPS Machine Learning for Audio Workshop, 2023.

\bibitem{GTZAN}
\BIBentryALTinterwordspacing
G.~Tzanetakis and P.~Cook, ``{Musical genre classification of audio signals},'' \emph{IEEE Transactions on Speech and Audio Processing}, vol.~10, no.~5, pp. 293--302, 2002. [Online]. Available: \url{https://doi.org/10.1109/TSA.2002.800560}
\BIBentrySTDinterwordspacing

\bibitem{MTT}
\BIBentryALTinterwordspacing
E.~Law, K.~West, M.~I. Mandel \emph{et~al.}, ``{Evaluation of Algorithms Using Games: The Case of Music Tagging},'' in \emph{Proceedings of the 10th International Society for Music Information Retrieval Conference, {ISMIR} 2009, Kobe International Conference Center, Kobe, Japan, October 26-30, 2009}.\hskip 1em plus 0.5em minus 0.4em\relax International Society for Music Information Retrieval, 2009, pp. 387--392. [Online]. Available: \url{http://ismir2009.ismir.net/proceedings/OS5-5.pdf}
\BIBentrySTDinterwordspacing

\bibitem{humphrey2018openmic}
E.~Humphrey, S.~Durand, and B.~McFee, ``Openmic-2018: An open data-set for multiple instrument recognition.'' in \emph{ISMIR}, 2018, pp. 438--444.

\bibitem{audioset}
J.~F. Gemmeke, D.~P.~W. Ellis, D.~Freedman \emph{et~al.}, ``Audio set: An ontology and human-labeled dataset for audio events,'' in \emph{Proc. IEEE ICASSP 2017}, New Orleans, LA, 2017.

\bibitem{wu2023large}
Y.~Wu, K.~Chen, T.~Zhang \emph{et~al.}, ``Large-scale contrastive language-audio pretraining with feature fusion and keyword-to-caption augmentation,'' in \emph{Proc. ICASSP}.\hskip 1em plus 0.5em minus 0.4em\relax IEEE, 2023, pp. 1--5.

\bibitem{chen2022hts}
K.~Chen, X.~Du, B.~Zhu \emph{et~al.}, ``Hts-at: A hierarchical token-semantic audio transformer for sound classification and detection,'' in \emph{ICASSP 2022-2022 IEEE International Conference on Acoustics, Speech and Signal Processing (ICASSP)}.\hskip 1em plus 0.5em minus 0.4em\relax IEEE, 2022, pp. 646--650.

\bibitem{liu2019roberta}
Y.~Liu, M.~Ott, N.~Goyal \emph{et~al.}, ``Roberta: A robustly optimized bert pretraining approach,'' \emph{arXiv preprint arXiv:1907.11692}, 2019.

\bibitem{park2019specaugment}
D.~S. Park, W.~Chan, Y.~Zhang \emph{et~al.}, ``Specaugment: A simple data augmentation method for automatic speech recognition,'' \emph{Interspeech 2019}, p. 2613, 2019.

\bibitem{SampleMatch}
\BIBentryALTinterwordspacing
S.~Lattner, ``Samplematch: Drum sample retrieval by musical context,'' in \emph{Proceedings of the 23rd International Society for Music Information Retrieval Conference, \{ISMIR\} 2022, Bengaluru, India, December 4-8, 2022}, 8 2022, pp. 781--788. [Online]. Available: \url{http://arxiv.org/abs/2208.01141 https://archives.ismir.net/ismir2022/paper/000094.pdf}
\BIBentrySTDinterwordspacing

\bibitem{koutini2022efficient}
K.~Koutini, J.~Schl{\"u}ter, H.~Eghbal-zadeh \emph{et~al.}, ``Efficient training of audio transformers with patchout,'' \emph{Interspeech 2022}, 2022.

\bibitem{castellon2021codified}
R.~Castellon, C.~Donahue, and P.~Liang, ``Codified audio language modeling learns useful representations for music information retrieval,'' \emph{International Society for Music Information Retrieval (ISMIR)}, 2021.

\bibitem{chen2022beats}
S.~Chen, Y.~Wu, C.~Wang \emph{et~al.}, ``Beats: audio pre-training with acoustic tokenizers,'' in \emph{Proceedings of the 40th International Conference on Machine Learning}, 2023, pp. 5178--5193.

\bibitem{nistal2024improving}
J.~Nistal, M.~Pasini, and S.~Lattner, ``Improving musical accompaniment co-creation via diffusion transformers,'' in \emph{Audio Imagination: NeurIPS 2024 Workshop AI-Driven Speech, Music, and Sound Generation}, 2024.

\bibitem{CyCLAP}
\BIBentryALTinterwordspacing
M.~Zhao, J.~Ono, Z.~Zhong \emph{et~al.}, ``{On the Language Encoder of Contrastive Cross-modal Models},'' oct 2023. [Online]. Available: \url{http://arxiv.org/abs/2310.13267}
\BIBentrySTDinterwordspacing

\bibitem{AudioFlamingo2}
\BIBentryALTinterwordspacing
S.~Ghosh, Z.~Kong, S.~Kumar \emph{et~al.}, ``{Audio Flamingo 2: An Audio-Language Model with Long-Audio Understanding and Expert Reasoning Abilities},'' mar 2025. [Online]. Available: \url{http://arxiv.org/abs/2503.03983}
\BIBentrySTDinterwordspacing

\bibitem{MaskCLIP}
\BIBentryALTinterwordspacing
X.~Dong, J.~Bao, Y.~Zheng, T.~Zhang, D.~Chen, H.~Yang, M.~Zeng, W.~Zhang, L.~Yuan, D.~Chen, F.~Wen, and N.~Yu, ``{MaskCLIP: Masked Self-Distillation Advances Contrastive Language-Image Pretraining},'' pp. 10\,995--11\,005, aug 2023. [Online]. Available: \url{https://arxiv.org/abs/2208.12262v1 http://arxiv.org/abs/2208.12262}
\BIBentrySTDinterwordspacing

\end{thebibliography}
